\documentclass[11pt]{article}
\usepackage{latexsym}
\usepackage{graphicx}
\usepackage{caption}
\usepackage{psfrag}
\usepackage{amsmath}
\usepackage{axodraw}
\usepackage{pstricks}
\usepackage{color}
\usepackage{booktabs}
\newcommand{\be}{\begin{equation}}
\newcommand{\ee}{\end{equation}}
\newcommand{\bea}{\begin{eqnarray}}
\newcommand{\eea}{\end{eqnarray}}

\newcommand{\nn}{\nonumber}
\newcommand{\lesssim}{ {\
\lower-1.2pt\vbox{\hbox{\rlap{$<$}\lower5pt\vbox{\hbox{$\sim$}}}}\ } }
\newcommand{\gtrsim}{ {\
\lower-1.2pt\vbox{\hbox{\rlap{$>$}\lower5pt\vbox{\hbox{$\sim$}}}}\ } }
\newcommand{\eg}{{\it e.g.}}
\newcommand{\ie}{{\it i.e.}}
\newcommand{\cf}{{\it cf.}}
\newcommand{\etc}{{\it etc.}}
\newcommand{\ansatz}{{\it ansatz}}

\input epsf

\numberwithin{equation}{section}

\begin{document}

\begin{titlepage}

\begin{flushright}
%\today
\end{flushright}
\vspace*{1.5cm}
\begin{center}
{\Large \bf Unraveling duality violations in hadronic tau decays}\\[2.0cm]

{\bf Oscar Cat\`{a},$^{a}$ Maarten Golterman$^{b}$ } and {\bf Santiago Peris$^{c}$}\\[1cm]

$^{a}$Ernest Orlando Lawrence Berkeley National Laboratory, University of California,\\
Berkeley, CA 94720, USA\\[.5 cm]
$^{b}$Department of Physics and Astronomy, San Francisco State
University,\\1600 Holloway Ave, San Francisco, CA 94132, USA\\[.5cm]
$^{c}$Grup de F{\'\i}sica Te{\`o}rica and IFAE\\ Universitat
Aut{\`o}noma de Barcelona, 08193 Barcelona, Spain\\[.5 cm]

\end{center}

\vspace*{1.0cm}

\begin{abstract}
There are some indications from recent determinations of the strong coupling constant $\alpha_s$ and the
gluon condensate that the Operator Product Expansion may not be accurate enough to describe non-perturbative
effects in hadronic tau decays. This breakdown of the Operator Product Expansion is usually referred to as
being due to ``Duality Violations.'' With the help of a physically motivated model, we investigate these
duality violations.  Based on this model, we argue  how they may introduce a non-negligible systematic error
in the current analysis, which employs finite-energy sum rules with pinched weights. In particular, this
systematic effect might affect the precision determination of $\alpha_s$ from tau decays. With a view to a
possible future application to real data, we present an alternative method for determining the OPE
coefficients that might help estimating, and possibly even reducing, this systematic error.
\end{abstract}

\end{titlepage}

%%%%%%%%%%%%%%%%%%%%%%%%%%%%%%
\section{Introduction}
\label{sec:intro}

Ever since the seminal work of Ref.~\cite{BNP}, it has been recognized that hadronic tau decays provide an
excellent experimental setting for extracting precise values of Standard Model parameters such as
the strong coupling constant $\alpha_s$ \cite{ALEPH}, the strange quark mass $m_s$ \cite{Davier}, and the CKM matrix element $|V_{us}|$ \cite{Gamiz:2004ar}.

The importance of a precise determination of QCD parameters can hardly be overemphasized, both in order to test the validity of the Standard Model as well as to constrain physics beyond the
Standard Model. In particular, $\alpha_s$ has a special status as the coupling
constant of QCD.
The current world average value as given by Ref.~\cite{Bethke:2006ac} is
\begin{equation}\label{worldalpha}
\alpha_s(M_z^2)=0.1189\pm 0.0010\ .
\end{equation}
Methods to determine $\alpha_s$ include, among others, deep inelastic scattering, the $e^+e^-$
cross section, the $Z$ width, and the lattice. What is  special  about the determination from tau decays is that it
extracts $\alpha_s$ at a remarkably low energy scale ($m_{\tau}=1.777$ GeV), but still high enough for
perturbation theory to be applicable and non-perturbative effects to be small.  Because  of asymptotic freedom,
the relative error on $\alpha_s$ gets reduced by roughly a factor of three after evolving it from the tau mass
 to the $Z$ mass. Recently, Ref.~\cite{Davier} performed an analysis based on
the ALEPH data collected at LEP, quoting as the final value
\begin{equation}\label{alephalpha}
\alpha_s(m_{\tau}^2)=0.345\pm 0.004_{\mathrm{exp}}\pm 0.009_{\mathrm{th}}\ ,
\end{equation}
which, extrapolated to the $Z$ mass, gives
\begin{eqnarray}\label{alphaZ}
\alpha_s(M_z^2)&=&0.1215\pm 0.0004_{\mathrm{exp}}
\pm 0.0010_{\mathrm{th}}\pm 0.0005_{\mathrm{evol}}\\
&=&0.1215\pm 0.0012\ .\nonumber
\end{eqnarray}
A somewhat more conservative estimate was given in Ref.~\cite{ALEPH}:
\begin{eqnarray}\label{aleph2}
\alpha_s(m_{\tau}^2)&=&0.340\pm 0.005_{\mathrm{exp}}\pm 0.014_{\mathrm{th}}\ ,\\
\alpha_s(M_z^2)&=&0.1209\pm 0.0006_{\mathrm{exp}}
\pm 0.0016_{\mathrm{th}}\pm 0.0005_{\mathrm{evol}}\nonumber\\
&=&0.1209\pm 0.0018\ .\nonumber
\end{eqnarray}
These results are compatible with the  OPAL value of $\alpha_s(M_z^2)=0.1219\pm 0.0010_{\mathrm{exp}}\pm
0.0017_{\mathrm{th}} $~\cite{OPAL} .

The results in (\ref{alphaZ}), together with the lattice result~\cite{Mason:2005zx}
\begin{equation}\label{latticealpha}
    \alpha_{s}(M_{z}^2)=0.1170\pm 0.0012\ ,
\end{equation}
are the most precise determinations of the strong coupling constant to date. Remarkably, they differ from
each other by 2.7 standard deviations. In fact, as emphasized by Bethke \cite{Bethke:2006ac}, these two
values extracted from tau decay and the lattice  differ from the world average computed with all the other
measurements by $+1.8$ and $-1.6$ standard deviations, respectively.  Given the relevance of these
measurements, we think it is timely to assess the validity of the  theoretical assumptions that go into the
estimates of systematic errors in both cases.  Our aim in this paper is to do this for the result extracted
from tau data.

We may subdivide the systematic uncertainties in the tau determination of the strong coupling according to three (not mutually independent) categories:
\begin{itemize}

\item[(i)] The truncation of the perturbation theory series in $\alpha_s$. This is sometimes considered to be the largest
uncertainty, and it is usually estimated by applying  different resummation techniques, such as fixed-order
perturbation theory (FOPT), contour-improved fixed-order perturbation theory (CIPT) or renormalon chain
resummation (RCPT).  This has been studied, \eg, in
Refs.~\cite{Davier,OPAL,CIPT2,CIPT,Neubert:1995gd,Jamin:2005ip}.

\item[(ii)] The truncation of the Operator Product Expansion (OPE). The OPE is
 an expansion in inverse powers of the momentum, each multiplied by a vacuum
condensate of increasingly higher dimension, which in any practical application
 has to be truncated at a certain order~\cite{SVZ,deRafael}.
Furthermore, the divergence of the perturbative series in $\alpha_s$ may be linked to the OPE through
renormalons~\cite{Zakharov:1992bx}. This suggests a strong correlation between perturbation theory and the
OPE. In particular, the gluon condensate  depends on the renormalization scheme and may also depend on the
order at which the series in $\alpha_s$ is truncated~\cite{Zakharov:2005cg}. However, if the same
prescription for the perturbative expansion is used in both vector and axial-vector channels, the gluon
condensate should also be the same in these two channels. The analysis of Refs.~\cite{ALEPH,Davier} currently finds
two different values for the gluon condensate from the vector and the axial-vector channels, given, \eg\  in CIPT, by
\begin{eqnarray} \label{gluonVA}
\frac{\alpha_s}{\pi}\langle GG\rangle\Big|_{\mathrm{Vector}}\!\!&=&(0.4\pm0.3)\times
10^{-2}\ {\mathrm{GeV}}^4\ ,\\
\frac{\alpha_s}{\pi}\langle
GG\rangle\Big|_{\mathrm{Axial}}&=&(-1.3\pm0.4)\times 10^{-2}\ {\mathrm{GeV}}^4\ ,\nn
\end{eqnarray}
which are incompatible within about 3 standard deviations. This suggests that the systematic errors may not
be fully understood. The situation concerning condensates of higher order is also unclear. See for example
the recent analyses in
Refs.~\cite{Bijnens:2001ps,Cirigliano:2003kc,Rojo:2004iq,Narison:2004vz,Friot:2004ba,Zyablyuk:2004iu,
Dominguez:2006ct,Almasy:2008xu}, and references therein.

\item[(iii)] The OPE is valid  in the euclidean region, where there are no
physical states. In contrast, on
the Minkowski axis the expansion is not guaranteed to work. This lack of convergence of
the OPE on the Minkwoski axis is usually referred to as due to
``Duality Violations'' (DVs). To our knowledge, the
best way to understand the problem
is at $N_c= \infty$, where the spectrum becomes an infinite set of poles. Even in this limit
the OPE has a cut due to the logarithms from the anomalous dimensions that survive in the
large-$N_c$ limit, and does not
reproduce  the set of infinite poles. Nevertheless,
the standard analysis of tau decays involves an integration
over a circle in the complex momentum plane which includes the Minkowski axis as well (see
Eq.~(\ref{cauchyope}) below). Of course, it is not known how much things might improve in the real-world
case, in which
$N_c=3$, relative to the limit $N_c\rightarrow \infty$,
 because at $N_c=3$ the infinite set of poles becomes a
cut. However, it is unlikely that the OPE will reproduce this cut systematically. Of course, this problem
has been widely recognized, and strategies have been introduced to ameliorate the problem.  One of those strategies
employed in both the ALEPH and OPAL analyses is the use of so-called ``pinched weights," in which zeroes in
the weight function suppress the region near the Minkowski axis~\cite{Le Diberder:1992fr} (see below).
While this  is expected to help, it is unknown to what degree, because there is no systematic theory
of DVs \cite{Shifman}. This makes it very difficult to estimate the associated systematic error.

\end{itemize}

In the method of analysis that has become standard for tau decays DVs are set to zero by fiat~\cite{BNP}. We
consider this situation to be unsatisfactory; if eventually DVs are to be dismissed, it should be as a
result of a data-driven analysis rather than as a result of not knowing how to take them into account. Our
aim, then, is to get a better understanding of the errors associated with DVs, as well as their interplay with the
use of the OPE. Lacking a theory of DVs that one could use to perform a systematic study in tau decays, the
best one can do is to resort to a model, and that is what we will do in this paper.

The advantage of a model is that one has full analytical control. Therefore, it can be used to test the
degree of accuracy obtained with the standard method of analysis for tau decays. The disadvantage is, of
course, that the model may not represent the situation in QCD. Although the usefulness of a model is thus
limited, we believe that a physically well motivated model can still give a fair idea of what to expect in
the real world. Compared to the standard analysis, in which DVs are not taken into account at all, our model
is in fact rather conservative, with DVs which are exponentially suppressed at higher energies. Other
reasonable models exist in which DVs are only power suppressed and which therefore may well produce larger
effects \cite{Shifman}.

%%%%%%%%%%%%%%%
%%%%%%%%%%%%%%%%%%%
\begin{figure}[t]
\begin{center}
\begin{picture}(209,174) (225,-120)
\SetWidth{0.5} \SetColor{Black} \Line(311,54)(311,-120) \SetWidth{0.9} \CArc(312,-33)(76.01,90.75,269.25)
\CArc(307.13,-27.55)(81.55,-87.28,-11.64) \ArrowArc(315.83,-29.28)(72.44,4.97,93.83)
\CArc(313.58,-32.5)(10.6,82.32,277.68) \SetWidth{0.5} \Line(225,-32)(317,-32) \Vertex(320,-32){2.81}
\ZigZag(322,-32)(392,-32){2}{10} \SetWidth{0.9} \Line(315,-43)(387,-43) \Line(315,-22)(388,-23)
\Text(404,-53)[lb]{\normalsize{\Black{${\mathrm{Re}}\,\, q^2$}}}
\end{picture}
\end{center}
\caption{\label{fig1}  Analytical structure of $\Pi(q^2)$ on the complex $q^2$ plane. The solid curve with
an arrow shows the contour to which Cauchy's theorem is applied.}
\end{figure}
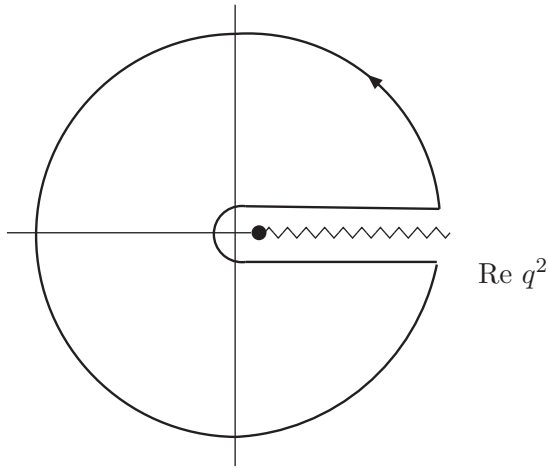

Let us start setting up the framework for our discussion. The main objects of study in hadronic tau decays
are the reconstructed spectral functions for the vector and axial hadronic channels. The vector and
axial current two-point correlators
$\Pi_{V,A}(q^2)$ are analytic everywhere in the complex $q^2$ plane except on the physical axis. Use of
Cauchy's theorem yields (see Fig.~\ref{fig1})~\cite{Shankar}
\begin{equation}\label{cauchy}
\int_0^{s_0}\,ds\,\, P(s) \,\frac{1}{\pi}{\mathrm{Im}}\,\Pi(s)=-\frac{1}{2\pi i}\oint_{|q^2|=s_0}\,dq^2
\,P(q^2)\, \Pi (q^2)\ ,
\end{equation}
where $P(q^2)$ is an arbitrary analytic function, and $\Pi$ is either $\Pi_V$ or $\Pi_A$.
As it stands, Eq.~(\ref{cauchy}) is, of course, exact.
However, if the radius $s_0$ is large enough, one may consider using the approximation $\Pi\approx
\Pi^{OPE}$  on the circle part of the contour.
Since, as we argued in point (iii) above, the OPE does not systematically reproduce the exact
function over the full circle, a correction term appears. We will denote this correction term by
${\cal{D}}^{[P]}$, resulting in the equation
\begin{equation}\label{cauchyope}
\int_0^{s_0}\,ds\,\, P(s)\, \frac{1}{\pi}{\mathrm{Im}}\,\Pi(s)=-\frac{1}{2\pi i}\oint_{|q^2|=s_0}\,dq^2
P(q^2) \Pi^{OPE} (q^2)+{\cal{D}}^{[P]}(s_0)  \ .
\end{equation}
The extra term ${\cal{D}}^{[P]}(s_0)$ represents the presence of duality violations. In all analyses of $\tau$ decay data to
date, ${\cal{D}}^{[P]}(s_0)$ has been assumed to vanish. In particular, the systematic error reported in
Eq.~(\ref{alephalpha}) only includes the convergence of perturbation theory (point (i) above).

One way to try to reduce the impact of non-perturbative  effects is by going to the highest possible energy
available, \ie,  $s_0=m_{\tau}^2$. Following this strategy, first put forward in Ref.~\cite{Le
Diberder:1992fr}, Refs.~\cite{ALEPH,OPAL} consider a family of spectral moments of the vector and
axial-vector correlators of the form
\begin{equation}
R_{kl}\propto  \int_0^{m_{\tau}^2}ds\ w_{kl}(s)\ \frac{1}{\pi}{\mathrm{Im}}\,\Pi(s)\ ,
\end{equation}
where the ``pinched weights'' $w_{kl}(s)$ are polynomials of the form
\begin{equation}\label{weights}
w_{kl}(s)= \left(1-\frac{s}{m_{\tau}^2}\right)^{2+k}\left(\frac{s}{m_{\tau}^2}\right)^{l}\
\left(1+2\frac{s}{m_{\tau}^2}\right)\ .
\end{equation}
The first factor vanishes for $s=m_\tau^2$ and therefore suppresses the contribution of duality violations
near the physical axis, where their impact is expected to be the largest. The second factor places more
weight on regions which are closer to the tau mass with increasing $l$, where perturbation theory should
work best. The third factor, as well as two powers of $(1-s/m_\tau^2)$ come from kinematics. A convenient
set of weights consists of the five combinations $w_{00}$, $w_{10}$, $w_{11}$, $w_{12}$ and $w_{13}$, which are
used to perform a combined fit to extract the four unknowns given by $\alpha_s$ and the condensates of
dimension $D=4,6$ and 8. Although these weights are sensitive to condensates up to dimension 16, both the DV
term $\mathcal{D}^{[kl]}(s_0) $ in Eq.~(\ref{cauchyope}) as well as the condensates of dimension higher than
8, have to be neglected in the fit in order to have less parameters than data points.

As we will see, in spite of questions about higher-order terms in the OPE and the possible presence of DVs,
the model we will present in Sec.~\ref{sec:model} shows that the results one obtains with the method of
pinched weights do fairly well in determining the central values for the coefficients of the OPE, including
the coefficient of the identity operator, which is the one used in the QCD analysis for extracting the value
of $\alpha_s$. However, this does not come about without an associated systematic error. In the model, this
systematic error turns out to be larger than the final error quoted in Eq.~(\ref{alephalpha}), suggesting
the possibility that the latter may have been underestimated. Moreover, if the model does not grossly
misrepresent the real case, it also shows that, in order to make a clear improvement in the reduction of
this systematic error, at least up to dimension-10 condensates as well as duality violations have to be
taken into account.

This paper is organized as follows. In Sec.~\ref{sec:theo} we summarize briefly the theoretical analysis of
hadronic tau decays. Section~\ref{sec:model} is devoted to introducing the model for the vector correlator.
We work out its high-energy expansion, and determine the duality violating terms analytically. In
Sec.~\ref{sec:aleph} we investigate the commonly-used ``moments method'' within our model. In
Sec.~\ref{sec:IIF} we propose an alternative method for the determination of the coefficients of the OPE,
which allows us to take higher-dimension operators and DVs into account, and we test this on our model,
comparing with the moments method.  We focus on the determination of the coefficient of the unit operator
because this is the one used for the extraction of $\alpha_s$ in the real-world case. Our conclusions and an
outlook for future directions are summarized in Sec.~\ref{sec:conc}.

%%%%%%%%%%%%%%%%%%%%%%%%%%%%%%
\section{Summary of theoretical framework for hadronic tau decays}
\label{sec:theo}

Extraction of experimental information on the inclusive hadronic tau decays to vector and axial-vector
channels proceeds through the analysis of the weak non-strange hadronic decays $\tau^-\rightarrow
(V^-/A^-)\nu_{\tau}$. It is convenient to work with the quantity
\begin{equation}
R_{\tau}^{V/A}\equiv \frac{\Gamma[\tau^-\rightarrow (V^-/A^-)\nu_{\tau}]}{\Gamma[\tau^-\rightarrow
\nu_{\tau}e^-{\bar{\nu}}_e]}\ ,
\end{equation}
in which the hadronic decay widths are normalized to electron decay. This quantity can be expressed
as~\cite{BNP}
\begin{equation}\label{spec}
R_{\tau}^{(V,A)}=12\pi
S_{EW}|V_{ud}|^2\int_0^{m_{\tau}^2}\frac{ds}{m_{\tau}^2}
\left(1-\frac{s}{m_{\tau}^2}\right)^2\left[\left(1+2\frac{s}{m_{\tau}^2}\right)
{\mathrm{Im}}\,\Pi_{V,A}^{(1)}(s)+{\mathrm{Im}}\,\Pi_{V,A}^{(0)}(s)\right]\ ,
\end{equation}
where $\Pi_{V,A}^{(J)}$ are the longitudinal ($J=0$) and transverse ($J=1$) components of the vector
and axial-vector vacuum polarization functions, defined by
\begin{eqnarray}\label{VV}
\Pi^{V,A}_{\mu\nu}(q)&=&i\int\mathrm{d}^{4}x\,
e^{iq\cdot x}\langle \,0\,|\,
T\lbrace\, J_{\mu}(x)\,J_{\nu}^{\dagger}(0)\,\rbrace |\,0\,\rangle\,\nonumber\\
&=&(q^{\mu}q^{\nu}-q^2g^{\mu\nu})\,\Pi_{V,A}^{(1)}(q^2)+q^{\mu}q^{\nu}\,\Pi_{V,A}^{(0)}(q^2) \ ,
\end{eqnarray}
in which $J_{\mu}^V(x)=\bar{u}(x)\gamma_{\mu}d(x)$ and $J_{\mu}^A(x)=\bar{u}(x)\gamma_{\mu}\gamma_5d(x)$.

For simplicity, we will consider only the vector correlator, in the isospin limit. This means that we set
${\mathrm{Im}}\,\Pi_{V}^{(0)}(s)=0$, rendering the vector correlator purely transverse. Furthermore, we will
also take the factor $S_{EW}|V_{ud}|^2\equiv 1$ since this factor can always be reinstated if needed. With
these changes, we define the moments $R_{kl}(m_{\tau}^2)$ as
\begin{equation}\label{momentsleft}
R_{kl}(m_{\tau}^2)=12\pi^2  \int_{0}^{m_{\tau}^2} \frac{ds}{m_{\tau}^2}\ w_{kl}(s)\
\frac{1}{\pi}\,\mathrm{Im}\,\Pi_{V}(s)\ ,
\end{equation}
which can also be re-expressed, with the help of Eq.~ (\ref{cauchyope}), as
\begin{equation}\label{momentsright}
R_{kl}(m_{\tau}^2)=6\pi i\, \oint_{|s|=m_{\tau}^2} \frac{ds}{m_{\tau}^2}\, w_{kl}(s)\
\Pi_V^{OPE}(s)+12\pi^2\,{\cal{D}}^{[kl]}(m_{\tau}^2)\ ,
\end{equation}
with $w_{kl}(s)$ given in Eq.~(\ref{weights}). In particular, $R_{00}(m_{\tau}^2)=R_{\tau}^{V}$ in
Eq.~(\ref{spec}).

%%%%%%%%%%%%%%%%%%%%%%%%%%%%%%
\section{The model}
\label{sec:model}

The model we will use to illustrate the possible role of DVs follows a suggestion of
Refs.~\cite{Shifman,Blok,Bigi}. It is a model for the vector two-point correlator which  contains an
infinite set of resonances with a finite width. Since DVs are a consequence of the analyticity properties of
the function $\Pi_V(q^2)$ in the complex $q^2$ plane, due care must be exercised to introduce this width in
a manner that will not spoil these properties.\footnote{In this regard, we note that a naive Breit-Wigner
expression is not good enough because it contains an imaginary part even for $q^2<0$ \cite{Shifman}.}
Specifically, the model is given by the expression
\begin{equation}\label{themodel}
    \Pi_V(q^2)=\frac{1}{\zeta}\left[\frac{2 F_{\rho}^2}{z+m^2_{\rho}}+
    \sum_{n=0}^{\infty}\frac{2 F^2}{z+m^2_V(n)}+ \mathrm{constant}\right]\ ,
\end{equation}
where
\begin{equation}\label{parameters}
z=\Lambda^2\left(\frac{-q^2-i\varepsilon}{\Lambda^2}\right)^{\zeta}\,\quad,\quad \zeta=1-\frac{a}{\pi
N_c}\quad , \quad  m_V^2(n)=m_0^2+n\ \Lambda^2\ .
\end{equation}
This model describes a separate rho meson together with an infinite tower of vector resonances with masses
on a Regge trajectory (a ``daughter trajectory'').
 Decay constants are given by
\begin{equation}\label{decayconstants}
    \langle\,0\,|\, V_{\mu}(0)\,|\rho_{n}(p,\lambda)\rangle =\sqrt{2}\ F \,m_V(n)\,\epsilon_{\mu}(p,\lambda)\ ,
\end{equation}
and widths by
\begin{equation}\label{widths}
    \Gamma_V(n)=\frac{a}{N_c}m_V(n)+{\cal{O}}\left(\frac{a^2}{N_c^2}\right)\ .
\end{equation}
Similar expressions hold for the rho meson. For illustration, factors of $N_c$ are inserted
into these expressions such as to allow for a comparison with the large $N_c$ limit of QCD.
However, in our analysis, we will take $N_c=3$. For a detailed discussion of properties of this type of
model, we refer to Refs.~\cite{CGP05,Golterman:2006gv}.

The function (\ref{themodel}) has a cut along the positive real axis in the $q^2$ plane and  satisfies the
once-subtracted dispersion relation
\begin{equation}\label{disp}
\Pi_{V}(q^{2})=q^2 \int_{0}^{\infty}\frac{dt}{t(t-q^{2}-i
\varepsilon)}\frac{1}{\pi}\,\mathrm{Im}\,\Pi_{V}(t)+\,\Pi_{V}(0)\ .
\end{equation}
The (infinite) subtraction constant $\Pi_{V}(0)$ has no imaginary part and will play no role in our
discussion since it does not contribute to equations such as (\ref{cauchy}).
We will neglect it in the following.

Using the relation
\begin{equation}
\lim_{N\rightarrow \infty}\
\sum_{n=1}^N\left(\frac{1}{n+z}-\frac{1}{n}\right)=-\psi(z)-\frac{1}{z}-\gamma_E\ ,
\end{equation}
where $\psi(z)=\Gamma\,'(z)/\Gamma(z)$ is the Euler digamma
function and $\gamma_E$ is the Euler-Mascheroni constant, it then follows that the full correlator can be
expressed in closed form:
\begin{equation}\label{green}
\Pi_{V}(q^2)=\frac{1}{\zeta}\left[\frac{2 F_{\rho}^2}{z+m_{\rho}^2}- \frac{2 F^2}{\Lambda^2}\
\psi\left(\frac{z+m_0^2}{\Lambda^2}\right)+ \mathrm{constant}\right]\ .
\end{equation}

\begin{figure}[t]
\centering %\psfrag{A}{$\mathrm{Re}(q^2)$}
\includegraphics[width=4.5in]{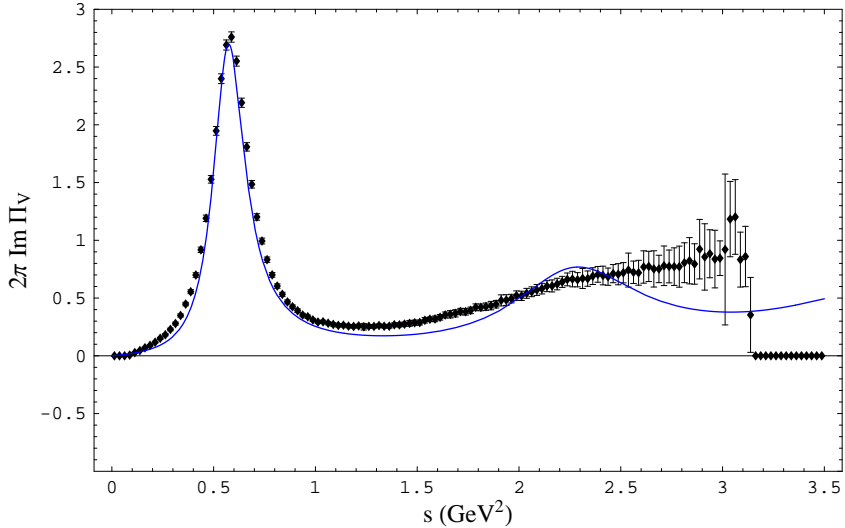}
\caption{Experimental data for the spectral function for the vector channel from the ALEPH
collaboration~\cite{ALEPH,Alephdata}. In blue, we show the curve given by the model defined in Eq.~(\ref{themodel})
with values for the parameters as given by Eq.~(\ref{param}).  }\label{fitspectrum}
\vspace{.5cm}
\end{figure}

Figure~\ref{fitspectrum} shows a comparison of the data from the ALEPH collaboration \cite{Alephdata} and the
imaginary part of $\Pi_V(q^2)$ in Eq.~(\ref{green}) with the following choice of values for the
parameters of the model:
\begin{equation}\label{param}
F_{\rho}=138.9\ \mathrm{MeV}\quad , \quad m_{\rho}=775.1\ \mathrm{MeV}\quad  , \quad F=133.8\
\mathrm{MeV}\quad ,\quad \frac{a}{N_c}=0.158\ .
\end{equation}
With these values, Eqs. (\ref{partonmodel},\ref{nodim2}) below determine $\Lambda$ and $m_0$ to be
\begin{equation}\label{param2}
    \Lambda=1.189\ \mathrm{GeV}\quad , \quad m_0=1.493\ \mathrm{GeV}\ .
\end{equation}
Although the model is clearly far from perfect, Fig.~\ref{fitspectrum} shows that it also does not
grossly misrepresent the main features of the spectrum, even though the value for the parameter $\zeta$ is
very close to unity ($\zeta\approx 0.95$). Therefore, the model illustrates how the large $N_c$ expansion
can be quite consistent with a realistic description of the real-world spectrum, for which $N_c=3$.

\subsection{Operator product expansion}
\label{sec:OPE}
Using the asymptotic expansion of the digamma function
\begin{equation}\label{asymp}
\psi(z)=\log{z}-\frac{1}{2z}-\sum_{n=1}^{\infty}\frac{B_{2n}}{2n\, z^{2n}} \quad , \quad |z|\gg 1\ ,\ -\pi<
\arg(z) < \pi\ ,
\end{equation} where $B_{2n}$ stand for the Bernoulli numbers
$ B_0=1 , B_1=-\ \frac{1}{2} ,  B_2=\frac{1}{6}$, \etc, it is straightforward to obtain the OPE for
the correlator (\ref{themodel}) as
\begin{equation}\label{opemodel}
\Pi_{V}^{OPE}(q^2) \approx -\ \frac{2 F^2}{\Lambda^2}\ C_0\
\log(-q^2)+\sum_{k=1}^{\infty}\frac{C_{2k}}{z^k}+ \mathrm{constant}
\end{equation}
where \cite{CGP05}
\begin{equation}\label{condensate}
C_0=1\quad ,\quad  C_{2k}=\frac{2}{\zeta} (-1)^k\left[  -F_{\rho}^2 \ m_{\rho}^{2k-2}+ \frac{1}{k}\
\Lambda^{2k-2}\ F^2\ B_k\left(\frac{m_0^2}{\Lambda^2}\right)\right] \ .
\end{equation}
The latter play the role of the vacuum condensates. In the previous expression, $B_k\left(x\right)$ stand
for the Bernoulli polynomials
\begin{equation}\label{Bpolynomials}
    B_n(x)=\sum_{k=0}^{n}\ \left(
                             \begin{array}{c}
                               n \\
                               k \\
                             \end{array}
                           \right)\ B_k\ x^{n-k}\ .
\end{equation}

Comparing with the case of QCD, we see that if we make the identification
\begin{equation}\label{partonmodel}
\frac{F^2}{\Lambda^2}=\frac{N_c}{24\pi^2}\ ,
\end{equation}
the coefficient of the parton model is reproduced. Furthermore, imposing the constraint
\begin{equation}\label{nodim2}
F_{\rho}^2=F^2\left(\frac{m_0^2}{\Lambda^2}-\frac{1}{2}\right)\ ,
\end{equation}
we require that the condensate of dimension two vanishes, \ie, $C_2=0$.\footnote{In QCD this would mean that we
neglect terms of order $m_q^2/z$ relative to those of
order $m_q\langle\overline{\psi}\psi\rangle/z^2$.}

Let us now look at the contribution from the OPE to the moments. Defining $R_{kl}^{(D)}$ as the contribution
of the dimension-$D$ operator to the moment $R_{kl}$, \ie, Eq.~(\ref{momentsright}) with
$\mathcal{D}^{[kl]}\equiv 0$, a straightforward calculation yields ($k\ge 1, j\geq 0$):
\begin{eqnarray}\label{momentstheworks}
R^{(0)}_{00}&=& \frac{12\pi^2 F^2}{\Lambda^2}\ C_0\ , \\
R^{(0)}_{1j}&=& \frac{144\pi^2 F^2}{\Lambda^2}\ C_0\ \frac{7+3 j}{(5 +j) (4+j) (3+j) (2+j) (1+j)}\ ,\nn \\
R_{00}^{(2k)}&=& -\ \frac{72\pi}{(m^{2}_{\tau})^{k\zeta}}\ C_{2k}\
\frac{\Lambda^{2k(\zeta-1)}\ \sin{k\pi\zeta}}{(k\zeta-4)(k\zeta-3)(k\zeta-1)} \ ,\nn \\
R_{1j}^{(2k)}&=& \frac{72\pi}{(m^{2}_{\tau})^{k\zeta}}\ C_{2k}\ \frac{\Lambda^{2k(\zeta-1)}\ (3k\zeta-7-3
j)\ \sin{k\pi\zeta}}{(k\zeta-5-j )(k\zeta-4-j)(k\zeta-3-j)(k\zeta-2-j)(k\zeta-1-j)} \  .\nn
\end{eqnarray}

In the limit $N_c\rightarrow \infty$ (\ie, $\zeta \rightarrow 1$), the contribution of the condensates to a
given moment is only non-vanishing whenever the dimension $2k$ is such that there is a zero in $k$ in the
denominator of the corresponding expression for the moment. The moments, however, are well defined for all
values of $k$. For instance, for $R_{00}^{(2k)}$, the only non-vanishing values happen for $k=1,3$ and 4,
which correspond to dimensions $D=2$, 6 and 8, in the limit $\zeta\to 1$. Since in the model $\zeta$ is very
close to one, these terms also dominate in the case of finite $N_c$.

\subsection{Duality violations}
\label{sec:dual} Up to now duality violations have not made an explicit appearance.
However, the restriction on the validity of the asymptotic expansion (\ref{asymp}) to the region $|\arg(z)|<
\pi$ implies that the OPE (\ref{opemodel}) cannot be valid on the full contour $|q^2|=s_0$, causing a
breakdown of the naive application of Cauchy's theorem  in which one would freely replace $\Pi(q^2)$ by
$\Pi^{OPE}(q^2)$ in Eq.~(\ref{cauchy}). The contribution from the Minkowski axis
($\arg(z)=\pi\leftrightarrow \mathrm{Re}\,q^2>0$) is responsible for this breakdown. This is how duality
violations make their appearance and why $\mathcal{D}^{[kl]}(m^2_{\tau})$ in Eq.~(\ref{momentsright}) does
not vanish in our model.

Use of the reflection property
\begin{equation}\label{reflection}
    \psi(z)=\psi(-z)-\pi\cot{(\pi z)}-\frac{1}{z}\ ,
\end{equation}
allows us to write an expansion for large values of $|q^2|$ which is valid for $\mathrm{Re}\,q^2>0$ (in
particular on the Minkowski axis):
\begin{equation}
\label{OPEdelta} \Pi_{V}(q^2)=\Pi_{V}^{OPE}(q^2)+\Delta(q^2)\ ,
\end{equation}
with the correction $\Delta$ given by
\begin{equation}\label{delta}
\Delta(q^2)=\frac{2 \pi F^2}{\Lambda^2}\frac{1}{\zeta}\left[-i+
\cot\left(\pi\left(\frac{-q^2}{\Lambda^2}\right)^{\zeta}+\pi\frac{m_0^2}{\Lambda^2}\right)\right]\ .
\end{equation}
This function $\Delta(q^2)$ behaves for large values of complex momentum $q^2$ as \cite{CGP05}
\begin{equation}\label{asymdelta}
\Delta(q^2)\sim e^{-2\pi\,\left(\frac{|q^2|}{\Lambda^2}\right)^{\zeta}|\sin\left\{(\pi-\phi)\zeta\right\}|},
\quad q^2=|q^2|e^{i\phi},\quad \phi \in [0,\pi/2]\cup[3\pi/2,2\pi)\ .
\end{equation}
This behavior shows that the limit $N_c\rightarrow \infty$ (\ie, $\zeta\rightarrow
1$) and the limit $\phi \rightarrow 0$ do not commute. If the limit $N_c\rightarrow \infty$ is taken first,
the expression (\ref{delta}) reveals the existence of poles at $\phi=0$ every time that
$q^2=m_0^2+ n \Lambda^2$.
This is as it should be, since in this case the spectrum of the model (\ref{themodel}) is composed of an
infinite set of Dirac deltas located precisely at these points. For finite $N_c$, on the Minkowski axis, one
obtains instead
\begin{eqnarray}
\frac{1}{\pi}\ {\mathrm{Im}}\,\Delta (t)&=& 2\ \frac{F^2}{ \zeta \Lambda^2 } \  \frac{\cos \left[2\pi
\left(\left( \frac{t}{\Lambda^2}\right)^\zeta \cos (\pi \zeta ) + \frac{m_0^2}{\Lambda^2}\right)\right]-
e^{-2\pi \sin[\pi \zeta ] \left( \frac{t}{\Lambda^2}\right)^\zeta }} {\cosh\left[2\pi
\left(\frac{t}{\Lambda^2}\right)^{\zeta}\sin(\pi \zeta ) \right]- \cos \left[2\pi \left(\left(
\frac{t}{\Lambda^2}\right)^\zeta \cos (\pi
\zeta) + \frac{m_0^2}{\Lambda^2}\right)\right]} \label{imdeltaexact}\\
&\approx & 4\ \frac{ F^2}{\zeta\Lambda^2}\ e^{-2\pi\left(\frac{t}{\Lambda^2}\right)^{\zeta}\sin(\pi\zeta)}\
\cos{\left[2\pi \left(\left(\frac{t}{\Lambda^2}\right)^{\zeta}\cos(\pi\zeta)+
\frac{m_0^2}{\Lambda^2}\right)\right] }\label{imdeltaasymp}\ ,
\end{eqnarray}
where the second expression is valid for $t$ large.

The DV function $\mathcal{D}^{[kl]}$, which corrects for the naive use of the OPE in Cauchy's
theorem as in (\ref{momentsright}), is given by (see Appendix and Ref.~\cite{CGP05})
\begin{equation}\label{DVimdelta}
{\cal{D}}^{[kl]}(m_{\tau}^2)=-\int_{m_{\tau}^2}^{\infty}ds \ w_{kl}(s)\ \frac{1}{\pi}\,{\mathrm{Im}}\,\Delta
(s)+ \mathcal{O}\left( e^{-\ 2\pi \frac{m^2_{\tau}}{\Lambda^2} }\right)\ .
\end{equation}
It follows from Eq.~(\ref{imdeltaasymp}) that the DV function  $\mathcal{D}^{[kl]}(m^2_{\tau})$ behaves like
$\sim e^{-2\pi \frac{m^2_{\tau}}{\Lambda^2}\ \frac{a}{N_c}}$. Therefore, duality violations are
exponentially suppressed in our model, and thus parametrically smaller than any term in the OPE.   However,
we also note that the exponential suppression is not that fast, because of the presence of the factor
$a/N_c\approx 0.158$ for the choice (\ref{param}).  Thus, DVs may be numerically important at or below the
tau mass, and the question is, by how much?

In order to answer this question, we would like to study to what extent one can determine the OPE
coefficients, given a certain analysis method.
To be more specific, consider our model, \cf\ Eq.~(\ref{opemodel}).  If one starts with a
general parametrization of the OPE of the form
\begin{equation}\label{opereal}
    \Pi_{V}^{OPE-\mathrm{fit}}(\lambda_{2k}; q^2) \approx -\ \frac{2 F^2}{\Lambda^2}\ \lambda_0\ \log(-q^2)+
    \sum_{k=1}^{\infty}\ \lambda_{2k}\ \frac{C_{2k}}{z^k}\ ,
\end{equation}
of course, the exact answer is given by $\Pi_{V}^{OPE-\mathrm{fit}}(\lambda_{2k}, q^2)$ with
$\lambda_0=\lambda_2=\lambda_4=\ldots=1$. How many of the parameters
$\lambda_{2k}, k=0,1,2,...$ can one  determine, and with what precision?%
\footnote{Since in our model we set the dimension-2 condensate
$C_2=0$, this means that the parameter $\lambda_2$ is irrelevant. We will disregard $\lambda_2$
in what follows.} What is a good method to minimize the error $|\lambda_{2k}-1|$ in this determination?

\section{The moments method applied to the model}
\label{sec:aleph} We are now in the position to test the standard method for determining
the OPE coefficients, \ie, the $\lambda's$ in Eq.~(\ref{opereal}).

As mentioned in the introduction, this method consists of a global fit to the
spectral moments $R_{\tau}=R_{00}$, $R_{10}$, $R_{11}$, $R_{12}$ and $R_{13}$, defined in Eq.~(\ref{momentsleft}),
evaluated at the tau mass, and taking into account only the contributions
from the OPE up to $D=8$. Of course, in the real world the parameter $\lambda_0$ is actually an
expansion in powers of $\alpha_s$ of the form
\begin{equation}\label{lambda0}
    \lambda_0=1+\frac{\alpha_s}{\pi}+ \ldots\ .
\end{equation}
Since our model is too simple to have any perturbative corrections, we will take the value obtained for
$|\lambda_0-1|$ as a rough estimate of the accuracy reached in determining $\alpha_s/\pi$.

Following Refs.~\cite{ALEPH,Davier,OPAL}, we compute the five spectral moments at the tau mass as a function of the
parameters $\lambda_{2k}$ with $k=0,2,3,4$, using Eq.~(\ref{opereal}), and making the replacement
$C_{2k}\rightarrow \lambda_{2k} C_{2k}$  in Eqs.~(\ref{momentstheworks}). The DVs, \ie, the functions
$\mathcal{D}^{[kl]}(m^{2}_{\tau})$ in Eq.~(\ref{momentsright}), as well as $\lambda_{2k}$ for $k>4$, are set
equal to zero. We then extract the values of  $\lambda_{2k}, k=0,2,3,4$ by doing a combined least-squares
fit to the five moments obtained from the ``data'' in Eq.~(\ref{momentsleft}), in which the true spectral
function $\frac{1}{\pi}\,\mathrm{Im}\,\Pi_V(q^2)$ of Eq.(\ref{green}), with parameters chosen as in
Eqs.~(\ref{param},\ref{param2}),
 is used.   Although within our model we have access to an infinite number of points,
 we have discretized the spectral function for the model in the same way as the ALEPH data have been
binned, \ie,
 into 125 bins with a binsize of $0.025$ GeV$^2$ each, with starting and ending points at
$s_0=0.0125$ GeV$^2$ and $s_0=3.1375$ GeV$^2$, respectively.

\begin{table}[t]
\centering
\begin{tabular}{cccccc}
\toprule%
& OPE$_\mathrm{dim-8}$ & OPE$_\mathrm{dim-8}$(DV) & IF$_\mathrm{dim-8}$ &
\textbf{IF$_{\mathrm{\textbf{dim-}}\mathbf{10}}$}& IF$_{\mathrm{dim-}10}$(no DV) \\%
\midrule $|\lambda_0-1|$ & 0.016 & 0.017 & 0.025 & $\mathbf{0.00077}$ & 0.0095 \\ [0.3ex] \midrule
$\lambda_4$ & 0.85 & 0.85 & 0.56 & $\mathbf{0.96}$ & 1.08  \\ [0.3ex] \midrule $\lambda_6$ & 0.65 & 0.65 &
0.56 &
$\mathbf{0.94}$ & 1.05 \\
[0.3ex] \midrule $\lambda_8$ & 0.53 & 0.54 & 0.31 & $\mathbf{0.84}$ & 0.97 \\ [0.3ex] \midrule
$\lambda_{10}$ & --- & --- & --- & $\mathbf{0.59}$ & 0.72 \\[0.3ex]
\bottomrule
\end{tabular}
\caption{Results of various fits discussed in Secs.~\ref{sec:aleph} and \ref{sec:IIF}.} \label{table1}
\end{table}

The results for the fit are shown in the second column of Table~\ref{table1}. As one can see, the values
obtained for the $\lambda's$ are fairly close to unity (in particular for the ``perturbative expansion''
$\lambda_0$), showing that the method works reasonably well. However, if the analogy (\ref{lambda0}) is
used, this translates into a systematic error in $\alpha_s$ of $15\%$, \ie, larger than the estimated
theoretical error shown in Eq.~(\ref{alephalpha}). Of course, this does not mean that the analysis of
Refs.~\cite{ALEPH,Davier,OPAL} must have a systematic error of this size. However, it does suggest that this method of
analysis is not safe against non-perturbative effects beyond those included in the ALEPH analysis.

Of course, in the context of our model, part of the reason may be that DVs have not been included, since the
functions $\mathcal{D}^{[kl]}(m^{2}_{\tau})$ are actually not zero. This is why we have repeated the
analysis with the same five
 moments and condensates, but now including the exact duality violations
of the model, \ie, Eq.~(\ref{DVimdelta}) calculated with the exact expression (\ref{imdeltaexact}).
In other words, we first ``subtracted'' the exact DVs from the ``data'' generated from the exact
spectral function before extracting the parameters $\lambda_{2k}$.  The
results are listed in the third column of Table \ref{table1}. As one can see, inclusion of duality
violations in the analysis of moments at $m_{\tau}$ does {\it not}
by itself lead to an appreciable change in the
results. Consequently,  we conclude that also the effect from higher dimension condensates may have to
be taken into account.

In fact, as already mentioned in the Introduction, the moment $R_{13}$ is sensitive to condensates with
dimension up to 16. The structure of the pinched weights (\ref{weights}), while being good at canceling the
integration region near the Minkowski axis, also makes the moments sensitive to condensates of higher
order, especially for larger $l$.
However, if we only use the moments at the fixed scale $m^2_{\tau}$,
there is simply no way to include more condensates while using the same moments
and still have a ``fit'' in which there are more data points than degrees of freedom.

%%%%%%%%%%%%%%%%%%%%%%%%%%%%%%
\section{An iterative fitting method}
\label{sec:IIF} The reason for using the values of the moments at the scale of the tau mass is that
non-perturbative effects are smaller at a higher scale, and that includes DVs as well.
However, in order to include the DV effects, we will have to go to lower scales to be able
to have more data points, as we will explain below. To this end, let us imagine that the tau mass is an
arbitrary scale we could change at our convenience, and let us refer to this scale as $s_0$. The moments are
now given by the same expressions (\ref{momentsleft}) and (\ref{momentsright}), but with the replacement
$m^2_{\tau}\rightarrow s_0$ everywhere.\footnote{This includes replacing  $m^2_{\tau}\rightarrow s_0$
in Eq.~(\ref{weights}).}  This trick has also been used by ALEPH and OPAL when studying
the validity of perturbation theory as a function of the scale $s_0$ \cite{ALEPH,OPAL}.

Based on experience with applying our  model to $\Pi_V-\Pi_A$, we propose a simple
parametrization of the DVs through the \ansatz\ \cite{CGP05}
\begin{equation}\label{sinexp}
\frac{1}{\pi}{\mathrm{Im}}\,\Delta^{DV-\mathrm{fit}}(s)=\kappa\, e^{-\gamma s}\sin{(\alpha +\beta s)}\ ,
\end{equation}
in which $\alpha, \beta, \kappa$ and $\gamma$ are parameters to be fitted. This \ansatz\ is of course
inspired by the asymptotic behavior of $\mathrm{Im}\,\Delta$ given in Eq.~(\ref{imdeltaasymp}).\footnote{In the
real QCD case, the function (\ref{sinexp}) will have to be compared to the experimental data.} We should
thus be able to fit it to our model to see if the results for the coefficients of the OPE improve relative
to the standard analysis described in the previous section.

Naively, one would think of fitting the moments to extract the values for all the OPE coefficients
$\lambda_{2k}$ together with the DV parameters $\alpha, \beta, \gamma$ and $\kappa$. However, this leads to
a complicated, multi-dimensional fit which yields poor and unstable results for the
parameters. Furthermore, the pinched weight moments have been designed \emph{not} to be very sensitive
to DVs, and thus  the determination of the DV parameters from the moments is difficult. On the other
hand, the spectral function is sensitive to DVs but less sensitive to the OPE coefficients
because they only enter through logarithms. Except for the parton model logarithm, these logarithms are
always screened: in the real world by an extra power of $\alpha_s$ and in the model by a factor $1-\zeta$.
Consequently, it appears to be best to combine using the spectral function for fitting the DVs with using
the moments for fitting the OPE coefficients.

Therefore, the strategy we propose consists of an iterative two-step procedure. Our starting \ansatz\ for the
spectral function is
\begin{equation}\label{fit}
\frac{1}{\pi}\,{\mathrm{Im}}\,\Pi^{\mathrm{fit}}_{V}(\lambda_{2k},s)=
\frac{1}{\pi}\,{\mathrm{Im}}\,\Pi_{V}^{OPE-\mathrm{fit}}(\lambda_{2k}; s) +
\frac{1}{\pi}\,{\mathrm{Im}}\,\Delta^{DV-\mathrm{fit}}(s) \ ,
\end{equation}
with the starting values $\lambda_{2k}=0, k\geq 1$, and the functions
$\Pi_{V}^{OPE-\mathrm{fit}}(\lambda_{2k}; s)$ and ${\mathrm{Im}}\,\Delta^{DV-\mathrm{fit}}(s)$ given by
(\ref{opereal}) and (\ref{sinexp}), respectively. Note that $\lambda_0$ is not set to zero in Eq.~(\ref{fit})
to account for the parton model contribution, which is the dominant part of the OPE and not
suppressed by $\alpha_s$ in the real world (or by $1-\zeta$ in our model).
 As before,
 we discretized the spectral function for the model
 into 125 bins with a binsize of $0.025$ GeV$^2$ each, with starting and ending points at
$s_0=0.0125$ GeV$^2$ and $s_0=3.1375$ GeV$^2$, respectively.

 In a first step, the \ansatz\ (\ref{fit}) is fitted to the  ``data'' given by the exact spectral
 function $\frac{1}{\pi}\,\mathrm{Im}\,\Pi_V(s)$  over the interval
 $1.4125\ \mathrm{GeV}^2\leq s\leq 3.1375\ \mathrm{GeV}^2$. Values for the
parameters $\lambda_0, \alpha, \beta, \kappa$ and $\gamma$ are obtained from this least-squares fit. Once
these values are known, the DV function $\mathcal{D}^{[kl]}(s_0)$ can be calculated from
Eq.~(\ref{DVimdelta}) with Eq.~(\ref{sinexp}), and then be incorporated into the moments $R_ {kl}(s_0)$ of Eq.~(\ref{momentsright}).

In a second step, the values $\lambda_{0,4,6,8,10}$ are then determined from
 a second least-squares fit to the $R_{kl}(s_0)$ moments, which include the duality violations
 determined in the first step, as a
function of the scale $s_0$ in the window $2.0125\ \mathrm{GeV}^2 \leq s_0 \leq 3.1375\ \mathrm{GeV}^2$.
Note that $\alpha$, $\beta$, $\gamma$ and $\kappa$ are held fixed in this step, while $\lambda_0$ is of
course allowed to vary. The  values obtained for the OPE parameters $\lambda_{2k}$, for $k=0,2,3,4,5$, are
then held fixed while the function (\ref{fit}) is re-fit again in order to find a new set of values for
$\alpha, \beta, \gamma$ and $\kappa$. This second step is iterated until stability is found. The choice of
the lower end in the above fitting windows is dictated by the stability of the result as well as by the
minimum value of the least-squares function. The upper end is fixed by the tau mass, as it would be in the
real QCD case.

\begin{table}[t]
\centering
\begin{tabular}{cccccccc}
\toprule%
$s_{min} $(GeV$^2)$  & least-squares & $|\lambda_0-1|$ & $\lambda_4$  & $\lambda_6$ & $\lambda_8$ & $\lambda_{10}$ \\
\midrule 1.0125 & 7.4E-5 & 0.01181 & 0.828 & 0.808 & 0.701 & 0.472\\ [0.3ex]
\midrule 1.2125 & 1.8E-5 & 0.00589 & 0.901 & 0.876 & 0.778 & 0.531  \\
[0.3ex] \midrule 1.4125 & 8.2E-7 & 0.00077 & 0.963 & 0.935 & 0.843 & 0.590\\ [0.3ex]
\midrule 1.6125 & 3.3E-6 & 0.00026 & 0.975 & 0.947 & 0.857 & 0.602 \\
[0.3ex] \midrule 1.8125 & 2.5E-6 & 0.00223 & 0.934 & 0.903 & 0.794 & 0.529\\ [0.3ex] \midrule
2.0125 & 2.6E-5 & 0.00246 & 0.920 & 0.886 & 0.762 & 0.485 \\
\bottomrule
\end{tabular}
\caption{Results of the iterative fit as a function of $s_{min}$; see text.} \label{table2}
\end{table}

Table \ref{table2} shows the result of the fit to the least-squares function constructed with the three
moments $R_{00}$, $R_{10}$ and $R_{11}$ as a function of the lower end in the fit of the function
(\ref{fit}) to the spectrum, together with the values of $\lambda_{0,4,6,8,10}$. As one can see, the
smallest value of the least-squares function is obtained for $s_{min}=1.4125$~GeV$^2$, which we therefore take  to
be the ``best'' determination.  The values of $\lambda_{0,4,6,8,10}$  for this fit are also shown  in the
(bold-faced) fifth column of Table \ref{table1}. Note that the values at $s_{min}=1.6125$~GeV$^2$ are actually
closer to the exact values in the model, but the least-squares function is larger than at $s_{min}=1.4125$~GeV$^2$.
The spread of the values in the parameters $\lambda_{0,4,6,8,10}$ between these two cases can be taken as a
rough indication of the systematic error of our iterative method.

For the fit in the second step, we have tested different combinations of spectral moments. The best
determination in terms of stability and the smallest value of the least-squares function occur for the fit
with the three moments $R_{00}$, $R_{10}$ and $R_{11}$. The fits gradually worsen if one uses more or less
moments. The lack of improvement when adding extra moments agrees with the fact that higher-order moments
are sensitive to higher orders in the OPE. Therefore, the inclusion of higher moments is disfavored because
of the OPE truncation, as we already anticipated. The  lack of improvement with too few moments is related
to the fact that in that case the OPE parameters are too correlated and not sufficiently constrained to be
determined by the fit. Therefore, the three-moment fit happens to be the optimal choice in the model.

\begin{figure}[t]
\centering
\includegraphics[width=4.1in]{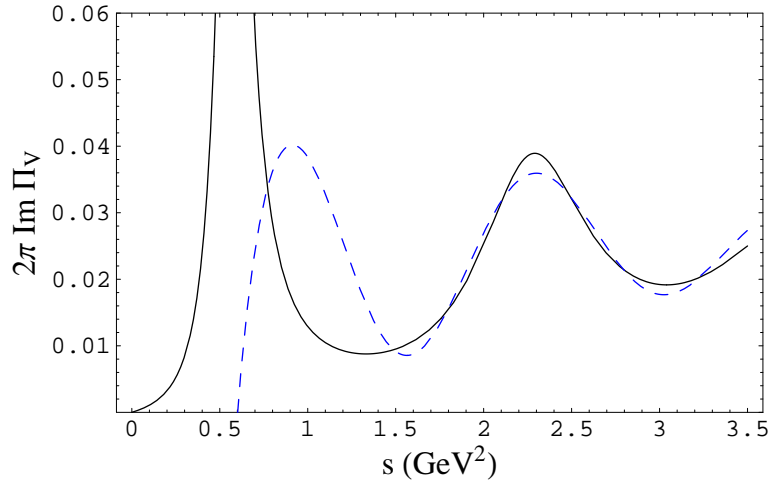}
\caption{Comparison between the exact spectral function (black solid curve) and the iterative fit described
in Sec.~\ref{sec:IIF} (blue dashed curve). }\label{fit1}
\end{figure}

Figure~\ref{fit1} shows the final estimate for the spectral function  Eq.~(\ref{fit}) obtained with the
parameter values from our iterative fit (blue dashed curve), in comparison with the exact spectral function (black solid curve).  We note that the rho is far away from the iterative fit.  This happens
because the rho is not part of the tower of resonances described by the digamma function in
Eq.~(\ref{green}); in other words, the rho is not part of the Regge trajectory.  This does not cause a
problem, since the approximation (\ref{sinexp}) is only needed for $s>s_0$, \cf\ Eq.~(\ref{DVimdelta}), where we always take
$s_0>2$~GeV$^2$.
The final values of the DV parameters extracted  from the fit are
\begin{equation}\label{dvparam}
\kappa=0.035,\qquad \gamma=0.51~\mathrm{GeV}^{-2},\qquad \alpha=12,\qquad \beta=-4.3~\mathrm{GeV}^{-2} \ .
\end{equation}
These numbers may be compared to the results obtained in the model, if we make the approximation $t^{\zeta}\approx
t$ in Eq.~(\ref{imdeltaasymp}). Since $\zeta\approx 0.95$, this approximation is justified. One
finds
\begin{equation}\label{truedvparam}
\kappa=0.051,\qquad \gamma=0.71~\mathrm{GeV}^{-2},\qquad \alpha=10,\qquad \beta=-4.5~\mathrm{GeV}^{-2} \ ,
\end{equation}
showing that the iterative fitting procedure does quite well for our model.

As already mentioned above,  for the determination of the OPE coefficients, \ie, the parameters
$\lambda_{2k}$, the result of the fit is summarized in the fifth column of Table~\ref{table1}. For
comparison, we also give in the fourth column the result we obtained if the condensates of dimension 10 and
higher are not included. Finally, the last column shows the results when dimension-10 condensates are
included, while at the same time duality violations are neglected (\ie, with $\kappa$ set equal to zero). In
this case, we note that the higher condensates actually move somewhat closer to their exact values. However,
the value obtained for $\lambda_0$, which is the parameter most directly related to $\alpha_s$, \cf\
Eq.~(\ref{lambda0}), is substantially worse.

Based on this table we draw the following conclusions, valid for our model. The inclusion of duality
violations is necessary in order to make a clear improvement in the determination of the different terms in
the OPE. However, this is only true provided enough of these OPE terms are included in the analysis as well.
In the case of the model, this requires the inclusion of up to dimension-10 condensates. We have checked
that including dimension-12 condensates in the analysis does not improve the determination of the parameters
$\lambda_{2k}$. However, the optimal number of terms in the OPE may depend on the case at hand, and does not
have to be the same in the case of QCD.

%%%%%%%%%%%%%%%%%%%%%%%%%%%%%%
\section{Conclusions and outlook}
\label{sec:conc}

The present analysis of data for hadronic tau decays is based on the use of finite-energy sum rules, in order to deal with the non-perturbative effects that cannot be neglected at a scale as low as the
tau mass.  This involves an extrapolation of the OPE
for the hadronic spectral functions over a full
circle around the origin in the complex momentum plane, including a region near the Minkowski axis where the physical spectrum is located
and the OPE is not expected to work. Although the contribution from this region is minimized by including
in the integrand convenient weight functions which vanish where the circle intersects the Minkowski axis,  contributions
from duality violations do not exactly vanish, and it is reasonable to expect that they have a
residual effect. This effect, if sizeable,  contaminates the determination of $\alpha_s$ and the
OPE condensates from hadronic tau decays.

In fact, there already seems to be some (perhaps circumstantial) evidence that there is something beyond the
OPE lurking in the tau data. As emphasized in the Introduction, the value of $\alpha_s$ extracted  from tau
decay is barely consistent with the other most precise determination, which is claimed by the
lattice. Moreover, the gluon condensate, which is the leading non-perturbative effect in the OPE, comes out
clearly different depending on whether the vector or the axial-vector channels are used in the analysis.

In spite of this, to the best of our knowledge, no analysis of duality violations has been done on the tau
data, and this gap has been the main motivation for the present work. With the help of a physically
motivated model, we have studied a method for extracting both the coefficient of the parton model logarithm
(from which $\alpha_s$ is extracted in the real case as a radiative correction),  as well as the OPE
condensates. For simplicity, we have taken the vector channel as the case study, but the axial-vector
channel can in principle be treated similarly. The obvious advantage of having a well-defined model is that
the exact answer is known. Nevertheless, we have tried to simulate a semi-realistic situation by
imagining fitting a general parametrization of duality violations to the model, as if it provided the real
``data.'' This allowed us to first test the standard method that has been used to date for analyzing tau
decay data. We have found that, taken at face value, this method does a fair job, in spite of the presence
of duality violations. The associated systematic error, however, would be too large to be acceptable in the
case of the real world, if the results obtained with our model are any indication. Needless to say,
we do not yet know if that is the case.

It is interesting to note that
our model provides a possible explanation for the fact that even if duality violations
are exponentially suppressed, they might still be numerically important at energy scales of order
a few GeV.  As discussed in Sec.~\ref{sec:dual}, the scale for this exponential decay involves
a factor $N_c$, and may thus differ substantially from the typical hadronic scale of 1~GeV.

In order to improve on the systematic error, we have proposed a new method of analysis based on an iterative
fit, described in Sec.~\ref{sec:IIF}. Our results show that a substantial improvement may be possible if one
is willing to do some modeling of the duality violations in the form of Eq.~(\ref{sinexp}).  At the same
time, our results indicate that also enough terms of the OPE should be taken into account.  In the model, we
needed to include condensates up to and including dimension ten. In the real world, this number of course
could be different.

Since there exists
at present no systematic theory of duality violations, we had to resort to a model and one
might thus dismiss the results on the basis that they are model dependent. However, we think that it is very
important to realize that, by the same argument, all the analyses of tau data performed so far have also
been done with a model, namely one in which duality violations are dismissed by fiat, and not by a fit to
the data. In terms of our \ansatz, Eq.~(\ref{sinexp}), current analyses are equivalent to setting the value
of $\kappa$ equal to zero.

Of course, it may turn out to be difficult to apply our new iterative fitting procedure to real-world data,
because these are afflicted with sizable and correlated errors which we completely ignored in the present
paper. In addition, in a fit over an interval such as proposed in Sec.~\ref{sec:IIF} applied to real data,
the running of $\alpha_s$ would have to be taken into account. However, again, that does not justify
ignoring duality violations altogether.  It would be interesting to see how our method fares in the case of
real data, even if it can only be used to establish bounds on the parameters $\kappa$ and $\gamma$.

\vspace{1cm}

\noindent \underline{Note added:} While this work was being finalized, two new determinations of
$\alpha_s$ at the tau mass appeared \cite{Baikov:2008jh,Davier:2008sk}, 
which, after running to the $M_Z$ scale,
translate into
\begin{eqnarray}
  \alpha_s(M_Z^2) &=& 0.1202\pm 0.0019\ , \\
  \alpha_s(M_Z^2) &=& 0.1212\pm 0.0011\ ,
\end{eqnarray}
respectively. These results differ from the lattice result 
(\cf\  Eq.~(\ref{latticealpha})) by 1.5 and 2.6 standard deviations,
respectively.

\vspace{1cm}

\section*{Acknowledgements}
%%%%%%%
We would like to thank M. Davier, M. Jamin, L. Mir, X. Prades, E. de Rafael and Z. Zhang for comments and
discussions. S.P. thanks the Department of Physics and Astronomy at San Francisco State University for its
warm hospitality. O.C. is supported by the Fulbright Program and the Spanish Ministry of Education and
Science under grant no. FU2005-0791, M.G. is supported in part by the US Department of Energy, and S.P. has
been supported by CICYT-FEDER-FPA2005-02211, SGR2005-00916, the Spanish Consolider-Ingenio 2010 Program CPAN
(CSD2007-00042) and by the EU Contract No. MRTN-CT-2006-035482, ``FLAVIAnet.''

\begin{appendix}

\section{Derivation of Eq.~(\ref{DVimdelta})}

\begin{figure}[t]
\begin{center}
\begin{picture}(209,174) (225,-120)
\SetWidth{0.5} \SetColor{Black} \Line(311,54)(311,-120) \SetWidth{0.9}
\DashCArc(307.13,-27.55)(82.55,-87.28,-11.64){2} \ArrowArc(306.13,-29.55)(45.55,-84.28,-17.64)
\DashCArc(313.83,-27.28)(74.44,4.97,91.83){2} \ArrowArc(310.83,-27.28)(39.55,7.08,90.64) \SetWidth{0.5}
\Line(225,-32)(317,-32) \Vertex(320,-32){2.81} \ZigZag(322,-32)(392,-32){2}{10} \SetWidth{0.9}
\ArrowLine(349,-43)(388,-43) \ArrowLine(388,-23)(350,-23) \ArrowLine(311,13)(311,47)
\ArrowLine(311,-110)(311,-74) \Text(404,-53)[lb]{\normalsize{\Black{${\mathrm{Re}}\,\, q^2$}}}
\Text(295,9)[lb]{\normalsize{\Black{$is_0$}}} \Text(294,45)[lb]{\normalsize{\Black{$i\infty$}}}
\Text(285,-115)[lb]{\normalsize{\Black{$-i\infty$}}} \Text(286,-77)[lb]{\normalsize{\Black{$-is_0$}}}
\end{picture}
\end{center}
\caption{ The contour used to evaluate Eq.~(\ref{fullD}), see text.}\label{fig3}
\end{figure}
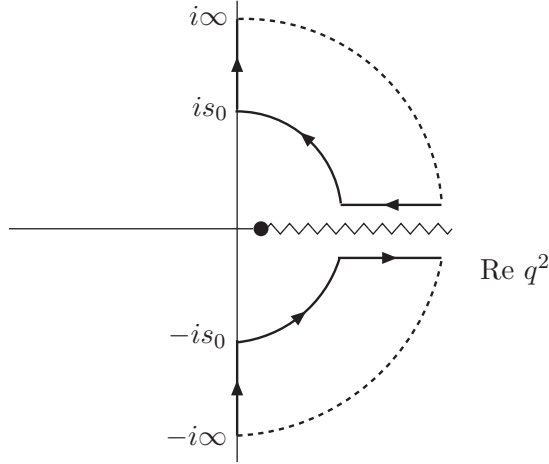

In Sec.~\ref{sec:theo} we already mentioned that the spectral moments can
be related to the large-$q^2$ expansion of the vector correlator as
\begin{equation}
R_{kl}(m_{\tau}^2)=6\pi i \oint_{|s_0|=m_{\tau}^2} \frac{ds}{m_{\tau}^2}\, w_{kl}(s)\,\Pi_{OPE}(s)+ 12 \pi^2
\, {\cal{D}}^{[kl]}(m_{\tau}^2)\ .
\end{equation}
The OPE part was discussed in Sec.~(\ref{sec:OPE}). The second term on the right-hand side corrects for
the fact that the OPE is not valid on the physical axis, and it can be expressed as
\begin{equation}\label{fullD}
{\cal{D}}^{[kl]}(s_0)= \int_{|q^2|=s_0\,,\
\mathrm{Re}\,q^2\geq 0}  \frac{dq^2}{m_{\tau}^2}\,w_{kl}(q^2)\,\Delta (q^2)\ ,
\end{equation}
where the contour runs counterclockwise only over the right half hemisphere, \cf\ the bold-faced semi-circle
in Fig.~\ref{fig3}. The function $\Delta(q^2)$ is defined in Eq.~(\ref{delta}).

Unlike in QCD, the model allows us to calculate
the duality-violating part as the difference between the full
correlator and its analytically-continued OPE expansion. Using
the reflection property of the Digamma function (\ref{reflection}), one
easily obtains Eq.~(\ref{delta}). In particular, note
that the resonance poles appear in $\Delta(q^2)$, as can be
seen through the identity
\begin{equation}
\pi\cot{(\pi z)}=\frac{1}{z}+2z\sum_{n=1}^{\infty}\frac{1}{z^2-n^2}\ .
\end{equation}
Expanding the cotangent for large $q^2$ and small $a/N_c$, one obtains Eq.~(\ref{asymdelta}). Note that for
small $a/N_c$, the damping factor along the imaginary axis ($\phi=\pi/2, 3\pi/2$) is stronger than along the
real axis ($\phi=0$):
\begin{eqnarray}\label{hier}
\Delta\left(\phi=\frac{\pi}{2},\frac{3\pi}{2}\right)&\sim& e^{-2\pi\,\frac{|q^2|}{\Lambda^2}}\ ,\nonumber\\
\Delta\left(\phi=0\right)&\sim& e^{-2\pi\,\frac{|q^2|}{\Lambda^2}\frac{a}{N_c}}\ .
\end{eqnarray}
Due to this difference, the leading contribution to the spectral moments can be easily derived by closing
the contour as shown in Fig.~\ref{fig3} and using Cauchy's theorem. Because of the exponential fall-off, the
integral vanishes at infinity, and we are left with integrals over the real and imaginary axes:
\begin{equation}
{\cal{D}}^{[kl]}(s_0)=-\int_{s_0}^{\infty}ds\,w_{kl}(s)\,
\frac{1}{\pi}{\mathrm{Im}}\,\Delta(s+i\epsilon)+\left\{\int_{-is_0}^{-i\infty}-
\int_{is_0}^{i\infty}\right\}dq^2 \,\omega_{kl}(q^2)\,\Delta(q^2)\ .
\end{equation}
Using the fact that the integral over the real axis dominates, \cf\ Eq.~(\ref{hier}), one finds the result
given in Eq.~(\ref{DVimdelta}).

\end{appendix}

%%%%%%%%%%%%%%%%%%%%%%%%%%%

\end{document}